\documentclass[traditabstract]{aa}

\usepackage{epsfig,graphicx,natbib}
\usepackage{amsmath,amssymb}

\def\nH2{\hbox{$n_\mathrm{H_2}$}}
\def\kms{\hbox{km\,s$^{-1}$}}
\def\PKS1830{\hbox{PKS\,1830$-$211}}
\def\B0218{\hbox{B\,0218$+$357}}
\def\cm-2{\hbox{cm$^{-2}$}}
\def\cm-3{\hbox{cm$^{-3}$}}
\def\fH2{\hbox{$f_{\rm H_2}$}}
 
\def\nH{\hbox{$n_{\rm H}$}}

\def\CLfive{\hbox{$^{35}{\rm Cl}$}}
\def\CLseven{\hbox{$^{37}{\rm Cl}$}}
\def\H2Cl+{\hbox{H$_2$Cl$^+$}}

\newcommand{\HH}  {\mbox{H$_2$}}       %  H2
\newcommand{\nCl}  {\mbox{Cl}}       % Cl
\newcommand{\nHCl}  {\mbox{HCl}}       % HCl
\newcommand{\OHp}  {\mbox{OH$^+$}}       %  OH+
\newcommand{\HHOp}  {\mbox{H$_2$O$^{+}$}}       %  H2O+
\newcommand{\HHClp}  {\mbox{H$_2$Cl$^+$}}       %  H2Cl+
\newcommand{\DDClp}  {\mbox{D$_2$Cl$^+$}}       %  D2Cl+
\DeclareSymbolFont{matha}{OML}{txmi}{m}{it} % txfonts
\DeclareMathSymbol{\varv}{\mathord}{matha}{118}

\begin{document}

\title{Chlorine-bearing molecules in molecular absorbers at intermediate redshifts\thanks{The spectra in Figure 1 are available in electronic form at the CDS via anonymous ftp to cdsarc.u-strasbg.fr (130.79.128.5) or via http://cdsweb.u-strasbg.fr/cgi-bin/qcat?J/A+A/}} 

\author{
S.\,H.\,J. Wallstr\"om \inst{1,2}
\and S.~Muller \inst{3}
\and E. Roueff \inst{4}
\and R. Le Gal \inst{5}
\and J.\,H.~Black \inst{3}
\and M.~G\'erin \inst{4}
}
\institute{Institute of Astronomy, KU Leuven, Celestijnenlaan 200D bus 2401, 3001 Leuven, Belgium
\and Institute of Astronomy and Astrophysics, Academia Sinica, 11F of Astronomy-Mathematics Building, No.1, Sec. 4, Roosevelt Rd., Taipei 10617, Taiwan
\and Department of Space, Earth and Environment, Chalmers University of Technology, Onsala Space Observatory, SE-43992 Onsala, Sweden
\and LERMA/LRA, Ecole Normale Sup\'erieure, Observatoire de Paris, CNRS UMR 8112, PSL Research University, Sorbonne Universit\'es, UPMC Universit\'e Paris, 24 rue Lhomond, 75005, Paris, France
\and Harvard-Smithsonian Center for Astrophysics, 60 Garden Street, Cambridge, MA 02138, USA
}

\date {Received / Accepted}

\titlerunning{Chlorine-bearing molecules in molecular absorbers}
\authorrunning{Wallstr\"om et al.}

\abstract{ We use observations of chlorine-bearing species in molecular absorbers at intermediate redshifts to investigate chemical properties and $^{35}$Cl/$^{37}$Cl isotopic ratios in the absorbing sightlines. Chloronium (H$_2$Cl$^+$) is detected along three independent lines of sight in the z=0.89 and z=0.68 molecular absorbers located in front of the lensed quasars \PKS1830\ and \B0218, respectively. Hydrogen chloride (HCl) was observed only toward \PKS1830, and is found to behave differently from H$_2$Cl$^+$. It is detected in one line of sight with an abundance ratio [H$_2$Cl$^+$]/[HCl] $\sim 1$, but remains undetected in the other, more diffuse, line of sight, with a ratio [H$_2$Cl$^+$]/[HCl]~$>17$. The absorption profiles of these two chlorine-bearing species are compared to other species and discussed in terms of the physical properties of the absorbing gas. Our findings are consistent with the picture emerging from chemical models where different species trace gas with different molecular hydrogen fraction. The $^{35}$Cl/$^{37}$Cl isotopic ratios are measured in the different lines of sight and are discussed in terms of stellar nucleosynthesis.

}
\maketitle

\section{Introduction}

Chlorine is a minor constituent of the Universe, with a solar abundance of only $\sim 3 \times10^{-7}$
relative to hydrogen \citep{Asplund2009}, that is orders of magnitude smaller than the elemental abundances of carbon, nitrogen, and oxygen. Nevertheless, chlorine has a strong chemical tendency to form hydrides (e.g., \H2Cl+, HCl) whose abundances can be as high as those of CH and H$_2$O, for example, in diffuse molecular clouds \citep{Neufeld2010, Sonnentrucker2010, Lis2010}. This is because of its unique thermo-chemistry properties: first, chlorine has an ionization potential of 12.97 eV, which is lower than that of hydrogen, such that it is predominantly in its ionized stage, Cl$^+$, in the atomic phase of the interstellar medium (ISM), and second, Cl$^+$ can react exothermically with H$_2$ triggering an active chemistry.

In fact, the interstellar chemical network of chlorine is thought to be relatively simple, dominated by a handful of hydrides \citep{Neufeld2009}, though the vast majority of interstellar chlorine will be in its neutral form in the presence of even small amounts of H$_2$. The reaction of Cl$^+$ with H$_2$ readily forms HCl$^+$, which has been detected in absorption in Galactic diffuse clouds \citep{DeLuca2012} harboring a significant amount (3--5\%) of the gas-phase chlorine. In turn, HCl$^+$ can further react exothermically with H$_2$ to form chloronium (\H2Cl+; first observed by \citealp{Lis2010}). Chloronium itself can react with free electrons to either form hydrogen chloride \citep[HCl; historically the first chlorine-bearing molecule detected in the ISM;][]{Blake1985} or release neutral chlorine. There are several pathways to destroy HCl through photodissociation, photoionization, or reactions with He$^+$, H$_3^+$, and C$^+$. 
Methyl chloride, CH$_3$Cl, was also recently identified in the young stellar object IRAS $16293-2422$ and the gaseous coma of comet 67P/Churyumov-Gerasimenko \citep{Fayolle2017}, suggesting that chlorine chemistry in space also extends to more complex species (see, e.g., \citealp{Acharyya2017} for gas-grain chemical models of chlorine chemistry).

From the nucleosynthetic point of view, elemental chlorine is produced on both short timescales in core-collapse supernovae, and long timescales in Asymptotic Giant Branch (AGB) stars and Type Ia supernovae \citep[p.164,][]{Clayton2003}. Chlorine is produced during oxygen burning, from fast reactions with the more abundant alpha elements, in stars massive enough to ignite it. In core-collapse supernovae excess neutrons cause a great increase in chlorine abundance, and hence they produce the majority of chlorine in the universe through explosive nucleosynthesis. Type Ia supernovae produce smaller amounts of chlorine through explosive oxygen burning.

Chlorine can be found in two stable isotopes, \CLfive\ and \CLseven, with different sources producing different relative abundances. Low metallicity core-collapse supernovae, like the ones that likely enriched the pre-Solar nebula, produce a \CLfive/\CLseven\ ratio of $\sim$3 as measured in the Solar system \citep{Asplund2009}.
Higher metallicity core-collapse supernovae can produce lower ratios \citep{Kobayashi2006,Kobayashi2011}. The isotopic ratios produced in AGB stars also depend heavily on metallicity, with low metallicities producing lower \CLfive/\CLseven\ values \citep{Cristallo2011}. 

Molecular absorbers at intermediate redshift ($z\sim1$) are powerful probes of the physico-chemical properties of distant galaxies. They offer a cosmological perspective on the chemical evolution of the Universe, for example with measurements of isotopic ratios at look-back times on the order of half or more of the present age of the Universe, but unfortunately, only a handful of such absorbers have been identified so far \citep[e.g.,][]{Combes2008,Wiklind2018}. 
In this paper, we collect observations of chlorine-bearing species in the two best studied and most molecule-rich redshifted radio molecular absorbers, with the goal of investigating their chemical properties and measuring the $^{35}$Cl/$^{37}$Cl isotopic ratio in different absorbing sightlines.

The molecular absorber located in front of the quasar \PKS1830\ is a face-on spiral galaxy at a redshift z=0.88582 (hereafter labeled MA0.89). The intervening galaxy lenses the quasar into two bright and compact images, leading to two independent sightlines in the southwest and northeast where absorption is detected (hereafter MA0.89~SW and MA0.89~NE), with impact parameters of $\sim$2 kpc and 4 kpc, respectively, in the absorber. These two compact, lensed images are embedded in a faint structure, reminiscent of an Einstein ring, seen at low radio frequencies \citep{Jauncey1991}.
A wealth of molecules and their rare isotopologs have been detected in this absorber, especially in MA0.89 SW, which is characterized by a large H$_2$ column density ($\sim 2 \times 10^{22}$~cm$^{-2}$) and moderate gas density $\sim 10^3$~cm$^{-3}$ \citep[e.g.][]{Wiklind1996, Muller2011, Muller2013, Muller2014}. MA0.89 NE is characterized by a lower H$_2$ column density ($\sim 1 \times 10^{21}$~cm$^{-2}$) and more diffuse gas composition as shown by the enhancement of the relative abundances of hydrides such as OH$^+$, H$_2$O$^+$, and H$_2$Cl$^+$ \citep[e.g.,][]{Muller2016b}.

The second molecular absorber (hereafter labeled MA0.68), located in front of the quasar \B0218, shares similar properties with MA0.89. It is a nearly face-on spiral galaxy at an intermediate redshift z=0.68466, also lensing the background quasar into two main compact images and an Einstein ring, seen at low radio frequencies and centered on the faintest image. Molecular absorption has only been detected in one line of sight (toward the brightest image), at an impact parameter of $\sim$2 kpc from the absorber's center \citep{Wiklind1995, Muller2007}. A number of molecules and their isotopologs have also been observed in MA0.68 \citep[e.g.][]{Wallstrom2016}, which has an H$_2$ column density ($\sim 8 \times 10^{21}$~cm$^{-2}$) that is intermediate between MA0.89 SW and MA0.89 NE.

With these two absorbers, we are thus exploring three independent lines of sight, with different redshifts, galactocentric distances, H$_2$ column densities, visual extinctions, metallicities, local radiation fields, cosmic-ray ionization rates, etc. Our observations and results are presented in Sections \ref{sec:obs} and \ref{sec:results}. We discuss our findings in terms of chemical properties derived from chlorine-bearing species and $^{35}$Cl/$^{37}$Cl isotopic ratios in Section~\ref{sec:discussion}.

\section{Observations} \label{sec:obs}

Observations of the two absorbers were obtained with the Atacama Large Millimeter/submillimeter Array (ALMA) between 2014 and 2017. A summary of these observations is given in Table~\ref{tab:obs}. Details of the observational setups and data analysis are described below for each absorber separately.

\begin{table*}[ht]
\caption{Summary of ALMA observations.}
\label{tab:obs}
\begin{center} \begin{tabular}{ccccccccc}
\hline \hline
Target & Species & Rest freq. $^{(a)}$ & Sky freq. $^{(b)}$ & $\delta v$ $^{(c)}$ & Dates of observations & Project ID \\
        &        &  (GHz)      &  (GHz) & (\kms)&              & \\
\hline
\PKS1830 & ortho-H$_2^{35}$Cl$^+$ & 189.225  & 100.341 & 1.5 & 2014, Jul. 21 and Aug. 26 & 2013.1.00020.S \\ 
         & ortho-H$_2^{37}$Cl$^+$ & 188.428  &  99.919 && observed simultaneously with previous &  \\ 

         & para-H$_2^{35}$Cl$^+$ & 485.418  & 257.404 & 1.1& 2014, Jun. 06 and Jul. 29 & 2013.1.00020.S \\    
         & para-H$_2^{37}$Cl$^+$ & 484.232  & 256.775 && observed simultaneously with previous & \\

         & H$^{35}$Cl & 625.919 & 331.908 & 0.9 & 2014, May 3, 6 & 2012.1.00056.S \\ 
         &            & &  && 2014 Jun 30 & 2013.1.00020.S \\
         & H$^{37}$Cl & 624.978 & 331.409 &&  observed simultaneously with previous &  \\   
\hline
\B0218 & ortho-H$_2^{35}$Cl$^+$ & 189.225 & 112.322 & 2.6 & 2016 Oct 22, 2017 May 02 & 2016.1.00031.S \\
       & ortho-H$_2^{37}$Cl$^+$ & 188.428 & 111.849 && observed simultaneously with previous & \\
\hline
\end{tabular} 
\tablefoot{
$a)$ Rest frequencies of the strongest hyperfine component, taken from the Cologne Database for Molecular Spectroscopy \citep[CDMS; ][]{Muller2001}.
$b)$ Sky frequencies are given taking redshifts $z_{abs}=0.88582$ and 0.68466 for MA0.89 and MA0.68, respectively, in the heliocentric frame.
$c)$ Velocity resolution.
}
\end{center} \end{table*}

\paragraph{MA0.68:} Observations of \H2Cl+ were obtained on 2016 October 22 and 2017 May 02. During the first run, the precipitable water vapor content was $\sim 2.4$~mm. The array was in a configuration where the longest baseline was 1.7~km. The bandpass calibrator was J\,0237+2848 ($\sim 2.2$~Jy at 100~GHz) and the gain calibrator was J\,0220+3241 ($\sim 0.2$~Jy at 100~GHz). During the second run, the precipitable water vapor content was $< 1$~mm. The array's longest baseline was 1.1~km. The bandpass calibrator was J0510+1800 ($\sim 2.6$~Jy at 100~GHz) and the gain calibrator J\,0205+3212 ($\sim 0.9$~Jy at 100~GHz).
The data calibration was done with the CASA package following standard procedures. We further improved the data quality with a step of self-calibration of the phase, using a model of two point-sources for MA0.68. When combined, the two datasets result in a synthesized beam of $\sim 1.0 \arcsec \times 0.6 \arcsec$ (P.A.$=-176^\circ$) at 100~GHz, slightly larger than the separation of $0.3\arcsec$ between the two lensed images of the quasar. However, we extracted the spectra along each line of sight by directly fitting each visibility set separately using the CASA-Python task UVMULTIFIT \citep{Marti-Vidal2014}, with a model of two point-sources. This is possible in the Fourier plane, thanks to the high signal-to-noise ratio of the data and the known and simple geometry of the source. The final spectrum was the weighted average of the individual spectra.

\paragraph{MA0.89:} We complemented previous observations of the ortho and para forms of \H2Cl+\ toward MA0.89 presented by \cite{Muller2014} and \cite{LeGal2017} with new observations of HCl obtained with ALMA during three observing runs: in 2014 May 03, 2014 May 06, and 2014 Jun 30. Both HCl isotopologs were observed in the same spectral window, and within the same receiver tuning as the para-H$_2$O$^+$ data presented by \cite{Muller2017}. The data reduction followed the same method as for MA0.68 data described above (self-calibration on the quasar, extraction of the spectra from visibility fitting, and weighted-averaged final spectrum).

\section{Results} \label{sec:results}

\begin{table*}[ht]
\caption{Summary of results and line-of-sight properties.}\label{tab:ncol}
\begin{center} \begin{tabular}{cccccccc}
\hline \hline

Sightline & $z_{\rm abs}$ & A$_{V,tot}$ & N(H$_2$) & N(H) &  N(H$_2$Cl$^+$) $^a$ &  N(HCl) $^a$ & $^{35}$Cl/$^{37}$Cl \\
                  &             & (mag)       & (cm$^{-2}$) &  (cm$^{-2}$) &  (cm$^{-2}$)       & (cm$^{-2}$) &  \\        
\hline
MA0.89 SW  & 0.88582 & 20  & $2 \times 10^{22}$ & $1 \times 10^{21}$ & $(2.3 \pm 0.3) \times 10^{13}$ & $(2.7 \pm 0.5) \times 10^{13}$ & $2.99 \pm 0.05$ $^b$ \\ 
             &         &     &                   & &                     &                     & $3.28 \pm 0.08$ $^c$  \\ 
MA0.89 NE  & 0.88582 &  2  & $1 \times 10^{21}$ & $2 \times 10^{21}$ & $(5.2 \pm 0.5) \times 10^{12}$   & $< 2.4 \times 10^{11}$ & $3.3 \pm 0.3$ $^b$ \\  
\hline
MA0.68    & 0.68466 &  8 & $8 \times 10^{21}$ & $4 \times 10^{20}$ & $(1.4 \pm 0.2) \times 10^{13}$   & --                  & $2.2 \pm 0.3$ $^b$ \\

\hline
\end{tabular}
\tablefoot{ $a)$ Total column density, including both $^{35}$Cl and $^{37}$Cl isotopologs, and ortho and para forms (assuming OPR=3) of \H2Cl+; $b)$ Calculated from H$_2$Cl$^+$ isotopologs; $c)$ Calculated from HCl isotopologs.}
\end{center} \end{table*}

\begin{figure*}[ht!] \begin{center}
\includegraphics[width=\textwidth]{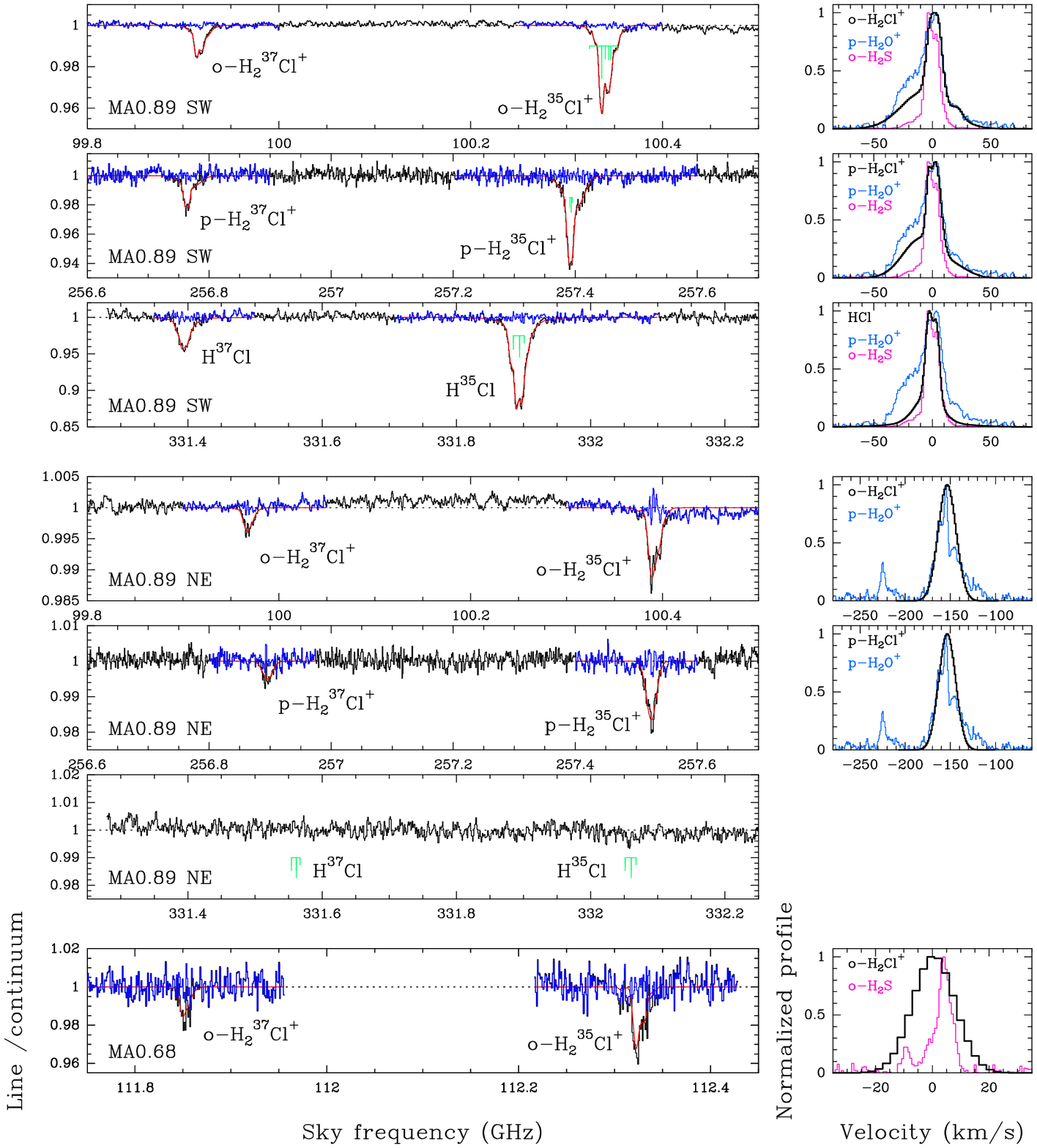}
\caption{Absorption spectra of chlorine-bearing species observed toward MA0.89~SW (top three panels), MA0.89~NE (middle three panels), and MA0.68 (bottom panel). Hyperfine structure is indicated in green for $^{35}$Cl-isotopologs of each line toward MA0.89~SW. Hyperfine structure was deconvolved by fitting multiple Gaussian velocity components; these fits are shown in red on top of spectra, with residuals shown in blue. On the right, deconvolved opacity profiles (assuming optically thin lines with $f_c=1$, as explained in Section~\ref{sec:results}), normalized to their peak opacity, are shown, together with opacity profiles of some other species for comparison. The profile of p-H$_2$O$^+$ is of the strongest hyperfine component of the $1_{10}$-$1_{01}$ transition, at rest frequency of 607~GHz. The profile of o-H$_2$S is the $1_{10}$-$1_{01}$ transition, at rest frequency of 168.8~GHz, which was observed simultaneously with o-\H2Cl+\ in the same tuning.}
\label{fig:spec}
\end{center} \end{figure*}

The spectra observed along the three lines of sight are shown in Fig.~\ref{fig:spec}. H$_2$Cl$^+$ is detected in all sightlines, unsurprisingly since it is ubiquitous in the Galactic diffuse and translucent gas \citep{Lis2010,Neufeld2012}. On the other hand, HCl is detected in only one of the two sightlines toward \PKS1830\ (MA0.89~SW), with a stringent upper limit toward the other. It was not observed toward MA0.68.

The \H2Cl+\ absorptions reach at most a few percent of the continuum level for all sightlines, so we must determine whether the absorption is optically thin and whether the source covering factor, $f_c$, is less than unity.
In MA0.89~SW, the ground state transition of species like CH$^+$ and H$_2$O is strongly saturated and reaches a depth of nearly 100\% of the continuum level, implying that $f_c \approx 1$ for these species \citep{Muller2014,Muller2017}. In MA0.89~NE, the situation is not so clear, but the deepest absorption reaches a depth of several tens of percent of the continuum level \citep{Muller2017}. For MA0.68, the covering factor of the 557~GHz water line is known to be large and close to unity \citep{Combes1997}, as confirmed by recent ALMA data (Muller, private communication).
The HCl absorption reaches a depth of $\sim$10\% of the continuum level toward MA0.89~SW, so without further evidence we consider $f_c=1$ for all species and all sightlines -- hence an optically thin line regime -- and our derived column densities (Table~\ref{tab:ncol}) can be viewed, strictly speaking, as lower limits.

When detected, both H$_2$Cl$^+$ and HCl show both their \CLfive- and \CLseven-isotopologs, observed simultaneously in the same spectral window (see a summary of the spectroscopic parameters of both species in Table~\ref{tab:hfs}). Considering optically thin lines and given the very small difference in their relative weight, the abundance ratio of the $^{35}$Cl- and $^{37}$Cl-isotopologs should reflect the $^{35}$Cl/$^{37}$Cl isotopic ratio (see Section~\ref{sec:35Cl/37Cl}).

In MA0.89, the ground-state transitions of both the ortho and para forms of \H2Cl+\ were observed. It should be noted that, from the same data, \cite{LeGal2017} determined ortho-to-para ratios (OPR) in agreement with the spin statistical weight 3:1 for both sightlines, within the uncertainties. In MA0.68, only ortho-\H2Cl+\ was observed. We adopted an OPR=3 for the calculations of \H2Cl+\ total column densities in all sightlines.

Finally, both \H2Cl+\ and HCl harbor a hyperfine structure that needs to be deconvolved before comparing intrinsic absorption profiles. This was done by fitting an intrinsic normalized profile, composed by the sum of individual Gaussian velocity components and convolved with the hyperfine structure (see Table~\ref{tab:hfs}), using $\chi^2$ minimization and taking the errors to be 1$\sigma$. Both the $^{35}$Cl- and $^{37}$Cl-isotopologs were treated in a common fit, with the $^{35}$Cl/$^{37}$Cl ratio as a free parameter. 
In order to assess whether the hyperfine structure deconvolution was robust, we fitted the ortho and para lines of \H2Cl+\ separately for MA0.89 SW. We obtained a good solution, with residuals at the noise level, with four Gaussian velocity components for this line of sight.
For MA0.89 NE and MA0.68 the detections have lower signal-to-noise ratios, so we limited the number of Gaussian components to one for sake of simplicity. These fits still yield residuals consistent with the noise level.
All the velocity profiles in Fig.~\ref{fig:spec} were normalized to the peak opacity to enable a straightforward comparison.

There is a striking difference in the opacity profiles of \H2Cl+\ and HCl toward MA0.89~SW: the former has prominent line wings, at velocities $| v | = 10-50$~\kms, while the latter is mostly a narrow component centered at $v=0$~\kms\ with FWHM $\sim 20$~\kms. In addition to the chlorine-bearing species of interest here, we also show in Fig.~\ref{fig:spec} the opacity profiles for some other species observed at the same epoch (to prevent effects from time variations of the absorption profile, see \citealp{Muller2008,Muller2014}). In MA0.89~SW, we find that the profile of \H2Cl+\ best matches that of H$_2$O$^+$, while the profile of HCl best matches that of H$_2$S. 
Similarly, in MA0.68, we observe that the FWHM of the \H2Cl+\ opacity is noticeably larger than that of H$_2$S, suggesting again that the two species trace different gas components.

Using the column density ratios between the two lines of sight of MA0.89, $\gamma_{SW/NE}$, \citet{Muller2016b,Muller2017} could classify the different species in two simple categories: those with low $\gamma_{SW/NE}$ ($\lesssim 5$), for example ArH$^+$, OH$^+$, H$_2$Cl$^+$, and H$_2$O$^+$, which are known to trace gas with low molecular hydrogen fraction, \fH2\footnote{ \fH2 = 2$\times$n(H$_2$)/[n(H)+2$\times$n(H$_2$)]},
of 1--10\% (see, e.g., \citealt{Neufeld2016}), and those with $\gamma_{SW/NE}$ above 20, for example CH, HF, HCO$^+$, HCN, and H$_2$S, tracing gas with much higher molecular hydrogen fraction, $>30$\% and up to 100\%. Since MA0.89~NE is dominated by gas with low \fH2\ the non-detection of HCl in this line of sight, with a large $\gamma_{SW/NE} > 90$, suggests that HCl is a tracer of high-\fH2 gas, in agreement with chemical predictions (see Section~\ref{sec:discussion}).
The line of sight towards MA0.68 is not yet fully characterized, but the estimated column density of H$_2$ and relative abundance of \H2Cl+\ suggest that it is intermediate between MA0.89~NE and SW in term of molecular hydrogen fraction.

\section{Discussion} \label{sec:discussion}

\subsection{Chemistry}

\begin{table*}[ht!]  
\caption{Abundance ratios and derived parameters for diffuse gas in MA0.89 SW and MA0.89 NE} 
\label{tab:relat}      
\begin{center}      
\begin{tabular}{c|c|c|c}    
\hline\hline       
       &    MA0.89 SW &  MA0.89 NE  & Reference/comment \\
\hline
[\HHClp] / [HCl] & 0.8 & $>$ 17  & present work \\
$[\rm{OH}^+]$ / [H$_2$O$^+$]  &  5.9  &  11  & \cite{Muller2016b} \\
\fH2\ & $\sim$ 0.04 & $\sim$ 0.02 & \cite{Muller2016b} \\
$\zeta$ (s$^{-1}$) & $2 \times 10^{-14}$ & $3 \times 10^{-15}$ & \cite{Muller2016b} \\
$x_e$ & $2.5 \times 10^{-4}$ & $2.4 \times 10^{-4}$ & using [OH$^+$]/[H$_2$O$^+$] (Eq.\ref{eq:xefromOH+/H2O+})  \\
G$_0$ & $\sim$ 0.5 $^\dagger$ & $>$ 10 & using [H$_2$Cl$^+$]/[HCl] and $x_e$ (Eq.\ref{eq:G0fromH2Cl+/HCl})  \\
\hline
\end{tabular}
\tablefoot{$\dagger$ Likely underestimated due to extinction.}
\end{center}
\end{table*}

Given the lack of observational constraints other than molecular abundances in the lines of sight through MA0.68 and MA0.89, and the fact that they sample different clouds and potentially different gas components, we have limited ourselves to a simple analytical model to explore the chlorine chemistry and how it varies with a few key parameters.
For diffuse cloud conditions, HCl and \HHClp\ are linked in a simple way, as HCl results from one specific channel of the dissociative recombination of \HHClp\ and is principally destroyed by photodissociation:

\begin{equation}
    \mathrm{\HHClp + e^- \rightarrow  HCl + H}
\end{equation}
\begin{equation}
    \mathrm{HCl + \gamma  \rightarrow   H + Cl}
\end{equation}
\noindent with reaction rates $k_1$ =  7.5 $\times$ 10$^{-8}$ (T/300)$^{-0.5}$ cm$^3$ s$^{-1}$ (see Appendix~\ref{app:ChemicalConsiderations}) and $k_2$ = 1.8 $\times 10^{-9}$ G$_0$ exp(-2.9 $A_V$) s$^{-1}$, where G$_0$ is the scaling factor of the interstellar UV radiation field (ISRF), expressed in Draine units \citep{Heays2017}. Under these conditions HCl is almost completely destroyed.

From these two reactions, we can write, at steady-state:
\begin{equation}
    [\HHClp ]/ [{\rm{HCl}}]  = \frac{k_2}{k_1 n_{e^-}}
\end{equation}

\noindent and, introducing the fractional ionization $x_e$ = $n(e^-) /n_{\rm{H}}$ and neglecting the $A_V$ dependence of the photodissociation rate as in a diffuse line of sight $A_V << 1$, we get

\begin{equation} \label{eq:G0fromH2Cl+/HCl}
    \frac{[\HHClp]}{[{\rm{HCl}}]} \sim 1.4 \times 10^{-2} \frac{G_0}{n_{\rm{H}} x_e} \left(\frac{T}{100}\right)^{0.5}.
\end{equation}
\noindent A large [\HHClp]/[HCl] ratio is obtained for large values of  G$_0$/$n_{\mathrm{H}}$ as suggested by \citet{Neufeld2009}.

We can perform a similar analysis for the [\OHp]/[\HHOp] ratio by considering the main formation process of \HHOp\ via the \OHp\ + \HH\ reaction and its destruction by dissociative recombination and reaction with molecular hydrogen:
\begin{equation}
    \mathrm{\OHp + \HH \rightarrow  \HHOp + H}
\end{equation}
\begin{equation}
    \mathrm{\HHOp + e^- \rightarrow  products}
\end{equation}
\begin{equation}
    \mathrm{\HHOp + \HH \rightarrow  H_3O^+ + H}
\end{equation}
\noindent with reaction rates $k_5 = 1.7 \times 10^{-9}$ cm$^3$ s$^{-1}$,  $k_6  = 4.3 \times 10^{-7}$ (T/300)$^{-0.5}$ cm$^3$ s$^{-1}$, and $k_7 = 6.4 \times 10^{-10}$ cm$^3$ s$^{-1}$.
Substituting the fractional ionization $x_e$ and the molecular fraction \fH2\ we get
\begin{equation*}
\frac{[\OHp]}{[\HHOp]} = \frac{k_6 [e^-] + k_7 [\HH]} {k_5 [\HH]}  = \frac{2 k_6 x_e + k_7 \fH2}{k_5 \fH2},
\end{equation*}
\begin{equation} \label{eq:xefromOH+/H2O+}
    \sim 8.8 \times 10^2 \frac{x_e}{\fH2}  \left(\frac{100}{T}\right)^{0.5} + 0.38 .
\end{equation}

Using the derived molecular fractions from \cite{Muller2016b}, we can estimate the fractional ionization from the [\OHp]/[\HHOp] ratio in the two MA0.89 lines of sight, assuming T = 100~K.
The [\HHClp]/[HCl] ratio then allows for the derivation of G$_0$, again following \cite{Muller2016b} in the assumption of $n_{\rm{H}}$ = 35 cm$^{-3}$.

The calculated values are reported in Table \ref{tab:relat}. As the derivation does not consider any dependence on the visual extinction, the derived values of the radiation field correspond to lower limits. 
We note that the SW line of sight has a significantly higher molecular fraction, and hence a larger total visual extinction ($A_V$; see Section~\ref{subsec:sightlines}), than the NE line of sight, so its ISRF is likely to be underestimated in our reasoning. The results depend strongly on the
value of $A_V$, but as an example: for a cloud of $A_V$ = 1 in MA0.89~SW, we calculate G$_0$ = 9.
This suggests that both MA0.89 SW and MA0.89 NE have more intense radiation fields than in the Solar neighborhood.

Our simple model is broadly consistent with the results of \citet{Neufeld2009}, though we note that the chlorine chemical network still has some uncertainties, as discussed in Appendix~\ref{app:ChemicalConsiderations}, which can significantly impact the model results.
\citeauthor{Neufeld2009} find a column density ratio [\HHClp]/[HCl]~$\sim1$, as we measure in MA0.89, for a two-sided slab model with G$_0$ = 10 and $n_{\rm{H}}$ = 315 cm$^{-3}$ at $A_V\sim1$. However, the predicted column densities of these molecules are significantly smaller than the observed $\sim$10$^{13}$~cm$^{-2}$ in our lines of sight.

\begin{figure}[ht] \begin{center}
\includegraphics[width=8.8cm]{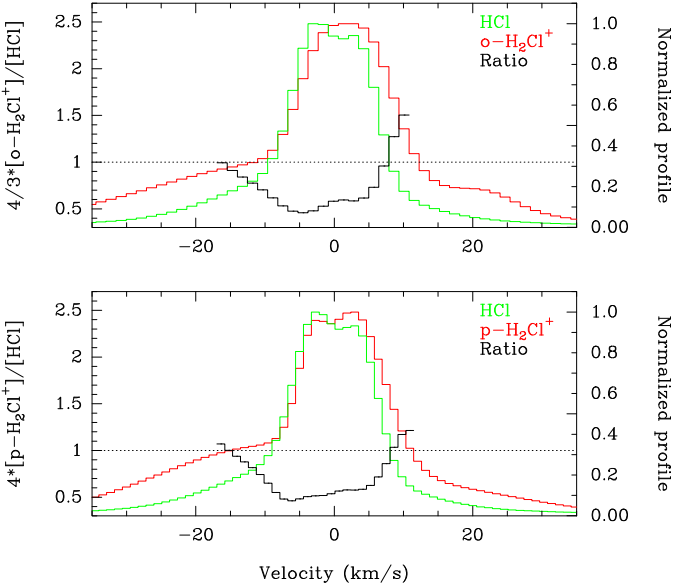}
\caption{Abundance ratio of H$_2$Cl$^+$ (derived from ortho and para forms, shown in top and bottom panels, respectively) over HCl across absorption profile toward MA0.89~SW.  
We use measured ortho-to-para ratio of three to obtain total H$_2$Cl$^+$ column density. Abundance ratio is calculated only for channels where signal-to-noise ratio of normalized profile of both \H2Cl+\ and HCl is higher than five.}
\label{fig:abundanceRatio-H2Cl+-HCl}
\end{center} \end{figure}

\subsubsection{Properties of each sightline} \label{subsec:sightlines}

Here we discuss more specifically our results towards MA0.68 and MA0.89 and compare the physical and chemical properties in the different sightlines. 

\paragraph{MA0.89 SW:}
This sightline is rich in molecular species, with almost 50 different species detected so far \citep[e.g.,][]{Muller2011,Muller2014}, and the H$_2$ column density is relatively high, at $\sim 10^{22}$~cm$^{-2}$. 
We can estimate the total visual extinction $A_{V,tot}$ in MA0.89 SW using Bohlin's law \citep{Bohlin1978} and the H and H$_2$ column densities \citep{Chengalur1999,Muller2014}. We find $A_{V,tot} \approx 20$ mag, which is shared among a number of individual clouds or kinematical structures. Obviously, there is at least one, most likely a few, cloud(s) with $A_V$ high enough ($\sim 1$) in the line of sight for HCl to be detectable. 
Indeed, the line profile of HCl indicates two peaks close to $v \sim 0$~\kms. Those two velocity components were previously noticed as two distinct clouds with different chemical setups, as they show opposite peaks in their CF$^+$ vs CH$_3$OH absorption \citep{Muller2016}, suggesting chemical segregation.

The multi-phase composition of the absorbing gas in this sightline has been clearly revealed by the recent observations of hydrides with ALMA \citep{Muller2014,Muller2016,Muller2017}.
Accordingly, the absorption from \H2Cl+\ and HCl can be simply understood
as tracing gas with different molecular fraction. In Fig.~\ref{fig:abundanceRatio-H2Cl+-HCl}, we see a clear difference between the core and wings of the line profiles: in the line center the gas is denser or better shielded and HCl dominates, with a [\H2Cl+]/[HCl] ratio of $\sim 0.5$, while in the wings the ratio increases to $\gtrsim 1$ and \H2Cl+\ dominates. This is consistent with different gas properties (i.e., \fH2) between the core and wings.

\paragraph{MA0.89 NE:}
Only \H2Cl+\ is detected in this line of sight, which is characterized by gas at low \fH2. 
The [H$_2$Cl$^+$]/[HCl] abundance ratio has a lower limit of $\sim$17, more than one order of magnitude higher than in the SW line of sight.
Following the method outlined above, we estimate a total $A_V \approx 2$ mag. Based on a statistical analysis, \cite{Muller2008} argued that the large time variations of the opacity profiles\footnote{due to the intrinsic activity of the background quasar, causing morphological changes, hence changing illumination to the foreground absorber.} imply a relatively small number of individual clouds in the line of sight, $\lesssim 5$, and that the bulk of the absorption should arise from clouds not much smaller than the background continuum source, $\sim 1$~pc. ALMA observations of the CH$^+$ absorption \citep{Muller2017} reveal a complex velocity profile which can be fitted with $\sim 10$ different Gaussian velocity components, some with widths as small as a few \kms, spanning a total velocity interval between $-300$~\kms\ and $-100$~\kms. Taking a simplified model of $\sim$10 individual clouds along the line of sight results in each having an average $A_V$ of $\sim$0.2. From the chemical predictions, we indeed expect some production of \H2Cl+\ at this level of $A_V$ but very little HCl, consistent with the non-detection of HCl.

\paragraph{MA0.68}
Here, we estimate a total $A_V \approx 8$ magnitudes. The overall absorption profile is narrow compared to that of MA0.89 and we estimate the number of clouds to be $\sim 3$ based on the number of distinct velocity features in the profile \citep[e.g.][]{Wallstrom2016}, suggesting characteristics intermediate between MA0.89~SW and MA0.89~NE. The abundance of \H2Cl+\ relative to H$_2$ is also intermediate between the MA0.89 sightlines. 

\subsubsection{Elemental chlorine and abundances of chlorine-bearing molecules}

Models of chlorine chemistry \citep[e.g.,][]{Neufeld2009} find that in the presence of H$_2$ most chlorine resides in the form of the neutral atom Cl. Observationally, a strong correlation is found between the column densities of Cl and H$_2$, in both the local ISM \citep{Jura1974,Sonnentrucker2006,Moomey2012} and in high-redshift damped Lyman-$\alpha$ absorption systems \citep{Balashev2015} over three orders of magnitude in H$_2$ column densities, with a relation\footnote{from an independent least-squares fit of the data from \citet{Moomey2012} and \citet{Balashev2015}.}

\begin{equation}
\mathrm{log[N_{Cl}]} = (0.79 \pm0.06) \times \mathrm{log[N_{H_2}]} - (2.13 \pm1.15) .
\label{eq:ClH2corr}
\end{equation}

\noindent This trend is found to be independent of the overall gas metallicity, suggesting that neutral chlorine could be an excellent tracer of H$_2$.

Assuming that the correlation in Eq.~\ref{eq:ClH2corr} also holds for our molecular absorbers at intermediate redshifts, we can infer the column density of neutral chlorine corresponding to the column density of H$_2$ in the different sightlines. 
With the observed column densities of H$_2$Cl$^+$ and HCl, we can then estimate the fraction of chlorine in these molecules (assuming neutral chlorine is vastly dominant and summing the \CLfive\ and \CLseven-isotopologs). Accordingly, we estimate that about 0.7\%, 1.8\%, and 0.9\% of chlorine is in the form of H$_2$Cl$^+$ for MA0.89~SW, MA0.89~NE, and MA0.68, respectively. For MA0.89~SW, we also estimate that about 0.9\% of Cl is in HCl.
These values are uncertain by about an order of magnitude (due to dispersion of the $\mathrm{N_{Cl}-N_{H_2}}$ correlation and uncertainty in the H$_2$ column density), and averaged over the whole sightline.

\subsubsection{HCl$^+$ and CH$_3$Cl} 

Besides \H2Cl+\ and HCl, we here discuss briefly some other chlorine-bearing species, already detected or potentially present in the ISM in our intermediate-redshift absorbers.
As previously mentioned, HCl$^+$ is the first chlorine-bearing species to be formed in the chlorine-chemistry network from the reaction of Cl$^+$ and H$_2$, in gas with a low enough \fH2 to maintain some Cl$^+$, and as such, HCl$^+$ is an excellent tracer of the diffuse gas component. HCl$^+$ has a complex spectrum, with multiple ground-state hyperfine transitions around 1.44~THz \citep{DeLuca2012}. Unfortunately, those transitions fall out of ALMA bands for a redshift of 0.89 and HCl$^+$ cannot be observed toward \PKS1830. It could in principle be observed toward \B0218 in ALMA Band~10, although with the observational challenge of a weak background continuum source at low elevation.

Another chlorine-bearing molecule, recently detected in the ISM, is methyl chloride (CH$_3$Cl). It has several hyperfine transitions which were covered in a deep spectral scan in the 7~mm band with the Australia Telescope Compact Array toward \PKS1830 \citep{Muller2011}. No features related to CH$_3$Cl (rest frequency near 79.8~GHz) are detected down to a few per mil of the continuum level.
Simulated absorption spectra for the conditions appropriate for the MA0.89 absorber yield a peak optical depth of $\tau = 1.95\times 10^{-3}$ for the hyperfine-structure blend, with an upper limit on the column density of $N({\rm CH}_3{\rm Cl}) = 5\times 10^{12}$ cm$^{-2}$. 
CH$_3$Cl was recently identified in the young stellar object IRAS $16293-2422$ by \citet{Fayolle2017}, who infer a CH$_3$Cl column density from these high-resolution ALMA data under the assumption of a fixed excitation temperature of 102 K of $4.6\times 10^{14}$ cm$^{-2}$ over a $0\farcs 5$ beam. A previous observation of HCl toward the same source by \citet{Peng2010} yielded well-constrained optical depths in the $J=1-0$ transition but referred to a much larger beam size of $13\farcs 5$. If the HCl result is re-interpreted for a fixed excitation temperature of 102 K, the resulting column density of HCl would be $8.9\times 10^{14}$ cm$^{-2}$, and if the ratio of column densities is simply scaled by the ratio of beam areas, then the abundance ratio would be [CH$_3$Cl]/[HCl] $= 7\times 10^{-4}$. This is likely to be an underestimate, because the HCl-emitting source probably does not fill the $13\farcs 5$ beam. Even so, the [CH$_3$Cl]/[HCl] ratio in IRAS $16293-2422$ is not inconsistent with the value $(4\pm 2)\times10^{-3}$ determined from the Rosetta Orbiter Spectrometer for Ion and Neutral Analysis (ROSINA) measurements of the gaseous coma of comet 67P/Churyumov-Gerasimenko \citep{Fayolle2017}. 
This [CH$_3$Cl]/[HCl] ratio is consistent with our upper limit on CH$_3$Cl in MA0.89~SW.

\subsection{$^{35}$Cl/$^{37}$Cl isotopic ratio} \label{sec:35Cl/37Cl}

The $^{35}$Cl/$^{37}$Cl isotopic ratio has been studied in a range of Galactic sources ranging from circumstellar envelopes to molecular clouds and star-forming regions. The measured ratios vary between $\sim$1 and 5, though most fall around 2.5 \citep{Maas2018} with uncertainties large enough to be consistent with the solar system ratio of 3.13 \citep{Lodders2009}. This value is measured from meteorites, and hence is indicative of the conditions at the formation of the solar system. Galactic chemical evolution models by \citet{Kobayashi2011} predict that the chlorine isotopic ratio in the solar system neighborhood has decreased, with increasing metallicity, since its formation to about 1.8 at the present day and Solar metallicity. 

From \H2Cl+\ absorption observed in 2012 in MA0.89~SW, \cite{Muller2014b} found a ratio $^{35}$Cl/$^{37}$Cl~=~$3.1 _{-0.2} ^{0.3}$, consistent with the Solar system value.
With the better quality of these new 2014 data, we revise this ratio ($2.99 \pm 0.05$) and we are also able to measure it in MA0.89~NE ($3.3 \pm 0.3$). These two measurements, at galactocentric radii of 2 and 4 kpc, respectively, are consistent each other, suggesting there are only small effects from metallicity or stellar population differences in the disk of MA0.89. In addition, we obtain another measurement from the HCl isotopologs in MA0.89~SW, $3.28 \pm 0.08$, relatively consistent with previous values and suggesting a good mixing between the different gas components with low and high \fH2.

In MA0.68, we find a $^{35}$Cl/$^{37}$Cl ratio of $2.2 \pm 0.3$, from H$_2$Cl$^+$ isotopologs. \citet{Wallstrom2016} found overall very similar C, N, O, and S isotopic ratios between MA0.68 and MA0.89~SW, which show clear evolution effects compared to the ratios found in the Solar neighborhood. The $^{35}$Cl/$^{37}$Cl ratio is the first to deviate between MA0.89~SW and MA0.68, and it would be interesting to confirm this difference with further observations.

A $^{35}$Cl/$^{37}$Cl ratio around three reflects nucleosynthesis products mainly from massive stars and core-collapse supernovae \citep{Kobayashi2011}. On the other hand, a \CLfive/\CLseven\ ratio of $\sim 2$ requires nucleosynthesis by either low-metallicity AGB stars or higher-metallicity Type II supernovae. As AGB stars affect their environment on long timescales, the first possibility requires MA0.68 to be a relatively old galaxy. For the second possibility, higher-metallicity Type II supernovae require short timescales and an intrinsically more metal-rich galaxy. The simplest explanation of the difference in chlorine isotopic ratios between MA0.89 and MA0.68 might be that MA0.68 has an intrinsically higher metallicity.

\section{Summary and conclusions} \label{sec:conclusions}

We investigate the absorption of chlorine-bearing molecules in two molecular absorbers (three independent lines of sight with different properties) at intermediate redshift, MA0.89 located toward \PKS1830\ ($z_{abs}=0.89$) and MA0.68 towards \B0218\ ($z_{abs} = 0.68$). \H2Cl+\ was observed, and detected, toward all sightlines. HCl was observed only toward \PKS1830, but detected only in one of the two sightlines.

Our results and conclusions are summarized as follows:
\begin{itemize}
\item The comparison of the absorption spectra of chlorine-bearing species between the different sightlines and with that of other species, namely H$_2$O$^+$ and H$_2$S, provides us with a simple classification based on molecular fraction (\fH2). \H2Cl+\ and HCl trace the gas component with low and high \fH2\ gas, respectively. This picture is consistent with predictions from chemical modeling, where HCl requires higher visual extinction ($A_V \sim 1$) than \H2Cl+\ to form.

\item These two chlorine-bearing species can hence be used together to characterize the column of absorbing gas. In particular, toward MA0.89~SW, we find that the [\H2Cl+]/[HCl] abundance ratio varies from $\sim$0.5 at the line center, to $\sim$2 in the line wings, reflecting the multi-phase composition of the gas along the line of sight.
The same ratio has a lower limit more than one order of magnitude higher toward MA0.89~NE.

\item The detection of the \CLfive- and \CLseven- isotopologs allows us to measure the \CLfive/\CLseven\ isotopic ratio at look-back times of about half the present age of the Universe. We find basically the same values at galactocentric radii of 2~kpc (\CLfive/\CLseven\ $= 2.99 \pm 0.05$ from \H2Cl+\ and $3.28 \pm 0.08$ from HCl) and 4~kpc (\CLfive/\CLseven\ $= 3.3 \pm 0.3$) in the disk of MA0.89 and in the Solar neighborhood (\CLfive/\CLseven\ $\sim 3$), while we find a lower ratio (\CLfive/\CLseven\ $= 2.2 \pm 0.3$) in MA0.68. This could be interpreted as MA0.68 having an intrinsically higher metallicity.
It would be interesting to obtain more measurements of the \CLfive/\CLseven\ isotopic ratio, including at higher redshifts, to investigate whether this isotopic ratio is a useful tracer of evolution.

\item The comparison of the observed abundances for \H2Cl+\ and HCl with that predicted from chemical modeling suggests the need for a stronger interstellar radiation field in the disk of these absorbers than in the Solar neighborhood. Evidence for a slightly increased cosmic-ray ionization rate of atomic hydrogen was also found in MA0.89 \citep{Muller2016b}, potentially related to the higher star formation activity at these intermediate redshifts.

\end{itemize}

In short, the two chlorine-bearing species \H2Cl+\ and HCl are sensible probes of the interstellar medium in redshifted absorbers. They offer diagnostics of the molecular fraction, gas composition, and nucleosynthesis enrichment via the \CLfive/\CLseven\ ratio. Their ground-state transitions are readily observable, for instance with ALMA, in a wide range of redshifts ($z \sim 0-6$), making them interesting tools to probe physico-chemical properties and evolution effects in the young Universe.

\begin{acknowledgement}
This paper uses the following ALMA data:\\
ADS/JAO.ALMA\#2012.1.00056.S \\
ADS/JAO.ALMA\#2013.1.00020.S \\
ADS/JAO.ALMA\#2016.1.00031.S. \\
ALMA is a partnership of ESO (representing its member states), NSF (USA) and NINS (Japan), together with NRC (Canada) and NSC and ASIAA (Taiwan) and KASI (Republic of Korea), in cooperation with the Republic of Chile. The Joint ALMA Observatory is operated by ESO, AUI/NRAO and NAOJ.

S.H.J.W. acknowledges support by the Ministry of Science and Technology of Taiwan under grants MOST104-2628-M-001-004-MY3 and MOST107-2119-M-001-031-MY3, and from Academia Sinica under AS-IA-106-M03.
\end{acknowledgement}

\bibliography{papers}
\bibliographystyle{aa.bst}

\newpage

\appendix

\section{Spectroscopic data}

\begin{table}[ht]
\begin{center} \begin{tabular}{cccccc}
\hline \hline
Species & \multicolumn{2}{c}{Quantum numbers} & Rest freq. & $S_{ul}$ & E$_\mathrm{low}$ \\
 & $J_{K_aK_c}$ & $F$ & (GHz) & & (K) \\
\hline
$o$-H$_2^{35}$Cl$^+$ & & & & & \\
 & 1$_{10}$--1$_{01}$ & 1/2--1/2 & 189.200369 & 0.50 & 0.0 \\
 & & 1/2--3/2 & 189.224382 & 2.50 & 0.0 \\
 & & 5/2--5/2 & 189.225063 & 6.30 & 0.0 \\
 & & 3/2--1/2 & 189.231910 & 2.50 & 0.0 \\
 & & 5/2--3/2 & 189.238598 & 2.70 & 0.0 \\
 & & 3/2--5/2 & 189.242471 & 2.70 & 0.0 \\
 & & 3/2--3/2 & 189.255994 & 0.80 & 0.0 \\
$o$-H$_2^{37}$Cl$^+$ & & & & & \\
 & 1$_{10}$--1$_{01}$ & 1/2--1/2 & 188.408840 & 0.50 & 0.0 \\
 & & 1/2--3/2 & 188.427815 & 2.50 & 0.0 \\
 & & 5/2--5/2 & 188.428370 & 6.30 & 0.0 \\
 & & 3/2--1/2 & 188.433716 & 2.50 & 0.0 \\
 & & 5/2--3/2 & 188.438900 & 2.70 & 0.0 \\
 & & 3/2--5/2 & 188.441923 & 2.70 & 0.0 \\
 & & 3/2--3/2 & 188.452800 & 0.80 & 0.0 \\
$p$-H$_2^{35}$Cl$^+$ & & & & & \\
 & 1$_{11}$--0$_{00}$ & 3/2--3/2 & 485.413427 & 1.33 & 0.0 \\
 & & 5/2--3/2 & 485.417670 & 2.00 & 0.0 \\
 & & 1/2--3/2 & 485.420796 & 0.67 & 0.0 \\
$p$-H$_2^{37}$Cl$^+$ & & & & & \\
 & 1$_{11}$--0$_{00}$ & 3/2--3/2 & 484.228484 & 1.33 & 0.0 \\
 & & 5/2--3/2 & 484.231804 & 2.00 & 0.0 \\
 & & 1/2--3/2 & 484.234298 & 0.67 & 0.0 \\
H$^{35}$Cl & & & & & \\
 & 1--0 & 3/2$_2$--3/2$_2$ & 625.901647 & 0.188 & 0.0 \\
 & & 3/2$_1$--3/2$_1$ & 625.901679 & 0.104 & 0.0 \\
 & & 5/2$_3$--3/2$_2$ & 625.918679 & 0.292 & 0.0 \\
 & & 5/2$_2$--3/2$_1$ & 625.918726 & 0.188 & 0.0 \\
 & & 1/2$_0$--3/2$_1$ & 625.931982 & 0.042 & 0.0 \\
 & & 1/2$_1$--3/2$_2$ & 625.932014 & 0.104 & 0.0 \\
H$^{37}$Cl & & & & & \\
 & 1--0 & 3/2$_2$--3/2$_2$ & 624.964357 & 0.188 & 0.0 \\
 & & 3/2$_1$--3/2$_1$ & 624.964387 & 0.104 & 0.0 \\
 & & 5/2$_3$--3/2$_2$ & 624.977782 & 0.292 & 0.0 \\
 & & 5/2$_2$--3/2$_1$ & 624.977830 & 0.188 & 0.0 \\
 & & 1/2$_0$--3/2$_1$ & 624.988257 & 0.042 & 0.0 \\
 & & 1/2$_1$--3/2$_2$ & 624.988289 & 0.104 & 0.0 \\
\hline
\end{tabular}
\end{center} 
\caption{Spectroscopic data for \H2Cl+\ and HCl. Data for \H2Cl+\ are from \citet{Araki2001}, taken from the Cologne Database for Molecular Spectroscopy \citep{Muller2001}. For HCl, entries are taken from \citet{Cazzoli2004}.} \label{tab:hfs}
\end{table}

\section{Considerations on the chemical network of chlorine-bearing species} \label{app:ChemicalConsiderations}

 \citet{Neufeld2009} have investigated the chlorine chemistry in different photodissociation region models where the thermal balance and chemical equilibrium are solved simultaneously for different densities, illumination conditions and visual magnitudes, and for two different values of the cosmic ionization rate: $1.8 \times 10^{-17}$~s$^{-1}$ and $1.8 \times 10^{-16}$~s$^{-1}$. That study allows for the derivation of different chlorine-chemistry regimes which affect the [\HHClp]/[HCl] ratio, where a large value is obtained for high radiation field and rather low density conditions. 

However, the dissociative recombination (DR) of \HHClp ~induces a major uncertainty in the chlorine chemical network, as discussed by \cite{Kawaguchi2016}, which severely impacts the abundances of \HHClp\ and HCl, and their ratio, as shown below. Three exothermic channels are available in the reaction:
\begin{align}
 \HHClp ~+ e^- & \rightarrow  \mbox{H} + \mbox{H} + \nCl \\
 \HHClp ~+ e^- & \rightarrow  \HH + \nCl \\
 \HHClp ~+ e^- & \rightarrow  \mbox{H} + \nHCl
\end{align}

The values of the different reaction rate coefficients used in the \cite{Neufeld2009} studies were unpublished results where the branching ratio towards HCl is taken as 10\% and the total DR rate coefficient is taken as $1.2 \times 10^{-7} \times (\frac{T}{300})^{-0.85}$~cm$^3$\,s$^{-1}$. That value was challenged by \cite{Kawaguchi2016} who report a total DR rate coefficient of $(0.46 \pm 0.05) \times 10^{-7}$ cm$^3$\,s$^{-1}$ at 209~K from infrared time-resolved spectroscopy experiments.
Last, but not least, \cite{Novotny2018} studied the dissociative recombination of the deuterated counterpart \DDClp\ in an ion storage ring experiment and obtained a total DR rate coefficient of $4.3 \times 10^{-7} \times (\frac{300}{T})^{0.5}$ cm$^3$\,s$^{-1}$. The derived coefficient is claimed to be most reliable at low temperatures relevant for astrochemical applications, in contrast to other experiments performed at room temperature. 
In addition, the branching ratio towards DCl was found to be 8.7\%. The dissociative recombination rate of \HHClp\ is expected to be larger according to previous experiments on other hydrogenated or deuterated molecular ions, which is also anticipated from theoretical considerations \citep{Jungen2009}. Indeed, \cite{Wiens2016} obtain a DR rate coefficient of $(1.1 \pm 0.3) \times 10^{-7}$~cm$^3$\,s$^{-1}$ for \DDClp\ and $(2.6 \pm 0.8) \times 10^{-7}$ cm$^3$\,s$^{-1}$ 
for \HHClp\ in an afterglow measurement at 300~K. \cite{Novotny2018} emphasize that their measurement of \DDClp~DR at 300~K is within the error bar of the storage ring experimental result but question the corresponding temperature extrapolation at low temperature. Summing up all these facts, we assume that the DR rate coefficient of \HHClp\ can be evaluated from the \DDClp\ measurement of \cite{Novotny2018}, assuming the same [\DDClp]/[\HHClp] ratio than that obtained by \cite{Wiens2016} at 300~K. The corresponding values are reported in Table \ref{tab:dr }.

\begin{table}[ht]
\caption{$\alpha$ factor corresponding to dissociative recombination rate of \HHClp\ expressed as $ \alpha \times (\frac{300}{T})^{0.5} $ cm$^3$\,s$^{-1}$.} 
\label{tab:dr }
\begin{center} \begin{tabular}{|l|c|c|c|}
  \hline
 full rate   & H  + H  + Cl  & Cl + H$_2$  & HCl + H \\
 \hline
$8.6 \times 10^{-7}$ & $7.4 \times 10^{-7}$  &  $4.3 \times 10^{-8}$  & $7.5 \times 10^{-8}$ \\
 \hline
\end{tabular} \end{center} \end{table}

These values correspond to a total destruction channel of \HHClp\ by electrons which is about seven times more efficient than that taken in \cite{Neufeld2009} at 300~K whereas the branching ratio to \nHCl\ is similar. The ratio between the two estimates reduces to a factor of four at 50~K and a factor of two at 10~K, due to the different slopes assumed for the energy dependence of the DR cross sections.
Given these uncertainties, a detailed chemical modeling appears superfluous.

\end{document}